# Epitaxial Graphene Growth on SiC Wafers

D.K. Gaskill[1], G.G. Jernigan[2], P.M. Campbell[2], J.L. Tedesco[1], J.C. Culbertson[2], B.L. VanMil[1], R.L. Myers-Ward[1], C.R. Eddy, Jr.[1] J. Moon[3], D. Curtis[3], M. Hu[3], D. Wong[3], C. McGuire[3], J.A. Robinson[4], M.A. Fanton[4], J.P. Stitt[4], T. Stitt[4], D. Snyder[4], X. Wang[4] and E. Frantz[4]

[1] Advanced SiC Epitaxial Research Laboratory, U.S. Naval Research Laboratory, Washington, DC 20375
[2] U.S. Naval Research Laboratory, Washington, DC 20375
[3] HRL Laboratories LLC, 3011 Malibu Canyon Road, Malibu, CA 90265
[4] The Pennsylvania State University, University Park, PA 16802

An *in vacuo* thermal desorption process has been accomplished to form epitaxial graphene (EG) on 4H- and 6H-SiC substrates using a commercial chemical vapor deposition reactor. Correlation of growth conditions and the morphology and electrical properties of EG are described. Raman spectra of EG on Si-face samples were dominated by monolayer thickness. This approach was used to grow EG on 50 mm SiC wafers that were subsequently fabricated into field effect transistors with $f_{max}$ of 14 GHz.

## Introduction

Graphene, a single sheet of $sp^2$ bonded carbon atoms possessing unique and unusual electrical, mechanical, physical, and chemical properties, has the scientific community searching for samples viable in new technological applications, such as post-CMOS digital electronics and high frequency (>100 GHz) analog electronics. This search is driving efforts to form epitaxial graphene (EG) on large area substrates. To that end, the formation of EG via the thermal desorption process of Si from SiC (graphenization) *in vacuo*, pioneered by Berger *et al.* (1) and demonstrated by others (2,3,4,5), is very attractive since SiC wafers up to 100 mm in diameter are commercially available.

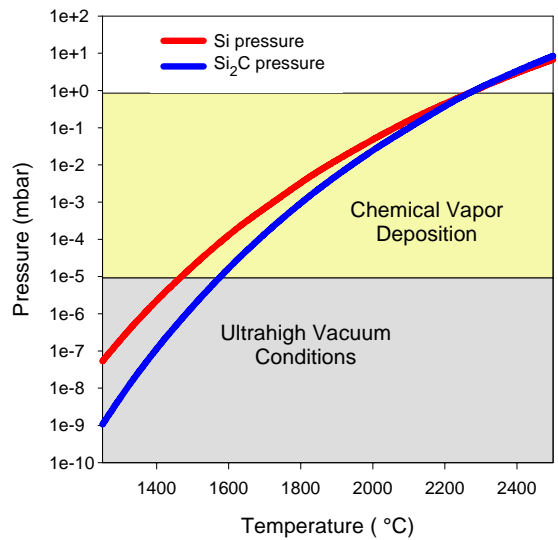

Figure 1. The vapor pressure of Si and $Si_2C$ over SiC as a function of temperature. The colored zones for chemical vapor deposition and ultrahigh vacuum operating conditions are shown. The curves were calculated from Ref. 6.

This *in vacuo* approach is conceptually simple as can be seen from Fig. 1. Here, the vapor pressure of Si over a SiC surface is plotted as a function of temperature (6). Thus to form EG the pressure over the substrate must be less than the Si partial pressure for a particular temperature. Here we describe the steps taken that resulted in EG grown on 50 mm SiC wafers using a commercial SiC chemical vapor deposition reactor.

## Experimental Details

Semi-insulating, on-axis (0°±0.5°), 50.8 mm diameter 4H- and 6H-SiC chemical-mechanical polished wafers were obtained from Cree (4H) and II-VI, Inc. (6H); both Si- and C-faces were acquired. The wafers were diced into 16×16 mm² samples prior to growth. *Ex-situ* chemical cleaning was performed on the SiC substrates prior to loading into an Aixtron/Epigress VP508 Hot-Wall CVD reactor. The samples were hydrogen etched at 100 mbar in 80 standard liters per minute of palladium purified hydrogen for 5 to 20 minutes at a temperature of 1600°C in order to remove polishing damage from the surface. Based upon prior work (7), we estimated that ~300 to 500 nm must be removed to eliminate surface polishing damage. Figure 2 is an example of a C-face hydrogen etched surface. Prior investigations in this laboratory (8) have shown that hydrogen etching results in a uniformly stepped, well-ordered surface for wafer diameters up to 76.2 mm for both C- and Si-face surfaces.

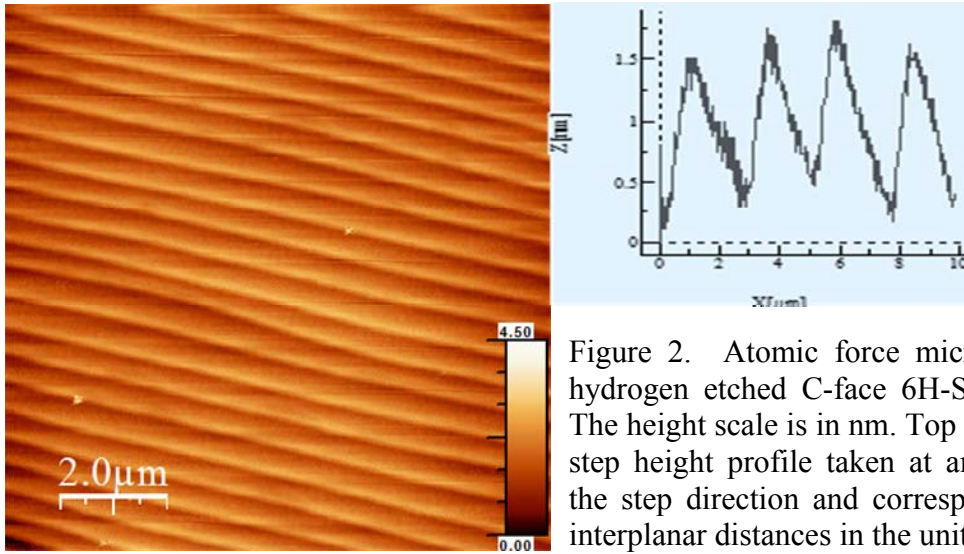

Figure 2. Atomic force micrograph of a hydrogen etched C-face 6H-SiC substrate. The height scale is in nm. Top right shows a step height profile taken at an angle from the step direction and corresponds to 6 Si interplanar distances in the unit cell.

Typically, the etched Si-face of both 4H and 6H polytypes are dominated by step heights of *ca.* 0.5 nm; this height corresponds to twice the Si interplanar distance in the SiC lattice and is often referred to as a bilayer step. The results on the C-face are different as the 4H (6H) polytype exhibits post-etch step height of 4 (6) interplanar distances. Following the hydrogen etch, the process chamber is purged with Ar and evacuated using the process pump (Ebara A25S). We have combined the hydrogen etch with graphenization into one process run (5) which is thought to yield a more pristine interface. A turbopump (Pfeiffer TMH 521) was used to achieve the high vacuum conditions necessary for EG formed in the temperature range of 1225 to 1700°C. The temperature uniformity across a 50.8 mm wafer during high vacuum graphenization is about 5°C. The pressure during graphenization, having durations of 10 to 120 minutes, began in the high $10^{-4}$ mbar range but steadily decreased over the course of the run, generally ending in the low $10^{-5}$ to high $10^{-6}$ mbar range.

Initially, the D, G and 2D Raman lines were used to confirm EG presence. Presently, EG was confirmed by measurable electrical resistance since semi-insulating substrates are used. The surface morphology of EG was characterized by Nomarski interference microscopy, tapping mode atomic force microscopy (AFM: Digital Instruments Dimension 3100), scanning electron microscopy (SEM: Carl Zeiss Surpra SS), scanning tunneling microscopy (STM: McAllister Technical Services) and field emission transmission electron microscopy (TEM: JEOL 2010F). After removal of backside EG growth via oxygen plasma etch, the sheet density and mobility were measured at both 300 K and 77 K using a van der Pauw configuration having Cu pressure contacts at the corners of the substrate. Measurement currents ranged from 1 to 100 µA and the magnetic field was 2,060 G. Some samples were processed for additional Hall measurements using standard photolithographic techniques to form patterns of crosses having 2 and 10 µm square active regions where electrical contact was made via Au pads. X-ray photoelectron spectroscopy (XPS) was performed using a monochromatic Al x-ray source (300 W), and C 1s spectra were acquired with a Omicron Sphera analyzer operating with a 20 eV pass energy and collected from a 630 µm diameter spot. A WITec confocal Raman microscope (CRM) with a 488 nm laser wavelength, diffraction limited lateral resolution of ~ 340 nm, and spectral resolution of 0.24 cm$^{-1}$ was utilized for Raman spectroscopy.

## Experimental Results

Our investigation initially focused on growing EG on both faces and polytypes of SiC using the 16×16 mm$^2$ samples. Substrate polytype was found to not impact EG properties. Yet, significant differences in EG on Si- and C-face samples were found. One difference was the observation of a transitional layer between the Si-face substrate and the EG; this transitional layer is absent for EG grown on C-face. Figure 3 shows this transitional layer as an additional peak at 285.6 eV in the C 1s XPS spectra. The C 1s peak characteristic of EG is located near 284.0 eV and the substrate peak is at 283.0 eV. The results of the XPS experiments are fully described in Ref. 9.

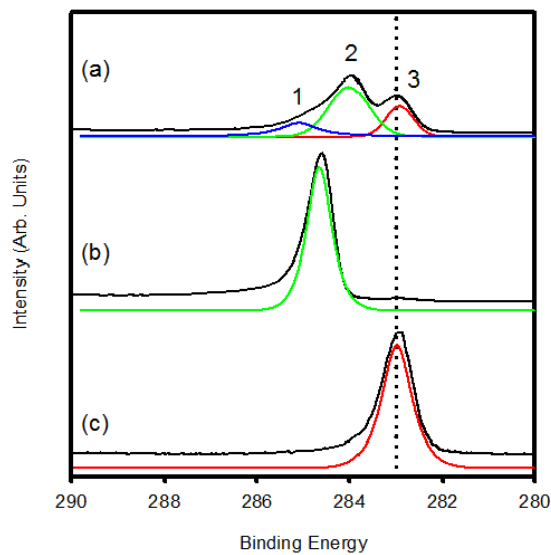

Figure 3. (color on-line) (a) The C 1s XPS spectra of EG on Si-face of 4H-SiC (highest line, in black). Fits to this data, and others in the figure, are below the highest lines. Peak 1 is the transitional layer, peak 2 is due to EG and peak 3 is due to the substrate. (b) EG on C-face shows the EG peak, which is shifted from the Si-face spectra due to charge screening. The substrate peak is small due to attenuation through the relatively think (~10 nm) EG. (c) An etched SiC substrate displaying only the substrate C 1s peak.

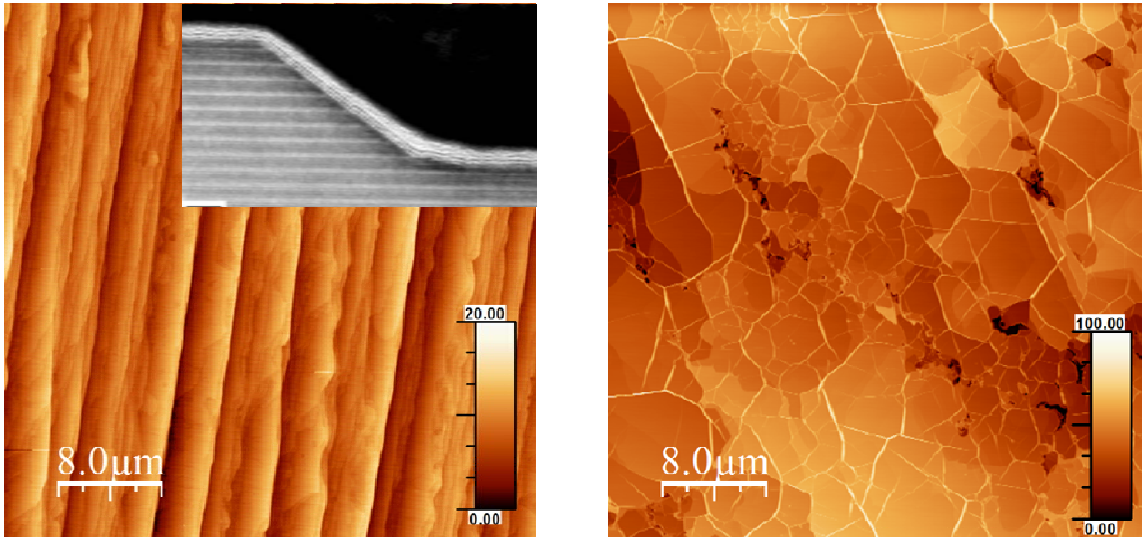

Figure 4. Atomic force micrograph (left) of EG grown on Si-face 6H-SiC. The inset (left) is a TEM cross section showing EG over a Si-face 6H-SiC step bunch. Atomic force micrograph (right) of EG grown on C-face 4H-SiC. A network of ridges ("giraffe stripes") can be discerned. The height scales are in nm.

The EG morphology on Si-face was markedly different from that of C-face EG. The morphology of the Si-face substrate underlying the EG is significantly rougher than that the pristine hydrogen etched surfaces, as shown by the step bunches in the AFM image of Fig. 4 (left). Yet, despite the increased step heights of the underlying substrate, EG readily "carpets" over the step bunches as seen in the TEM image in Fig. 4 (inset, left). In contrast, the C-face appears to be covered by a dense array of ridges ("giraffe stripes") that are up to 100 nm in height; see Fig. 4 (right). Furthermore, AFM and STM images show the giraffe stripes can be found in a wide array of orientations with respect to the step edges. Preliminary investigations at this laboratory suggest the ridges are out-of-plane graphene structures. A few pits can also be observed in Fig. 4 (right).

Growth conditions that resulted in high transport mobilities for Si-face EG samples were high temperatures, > 1500°C, and long times, ≥60 minutes. Growth conditions that yielded good transport values for C-face EG samples were low temperatures, ≤ 1500°C, and short times, ≤60 minutes. Under these growth conditions, C-face EG samples were thicker (3 to 23 nm) than Si-face EG samples (0.5 to 3.5 nm) grown simultaneously. The best transport results obtained were markedly different between the two faces. The highest 300 K mobilities, obtained for small patterned Hall crosses (2×2 or 10×10 μm$^2$, examples are shown in Fig. 5), were for Si-face: 1120 cm$^2$ V$^{-1}$ s$^{-1}$ having a hole sheet density of $8.5 \times 10^{11}$ cm$^{-2}$, and for C-face: 18,100 cm$^2$ V$^{-1}$ s$^{-1}$ having a hole sheet density of $2.1 \times 10^{12}$ cm$^{-2}$. These mobility values are comparable to the best reports for EG (1, 10).

Characterization of EG using Raman spectroscopy requires fitting the 2D Raman peak (11,12,13,14). Raman spectra of EG on the Si-face fit by one or four Lorentzian functions are characteristic of monolayer or bilayers EG, respectively (11), whereas a fit with two Lorentizians is indicative of bulk graphite; this is illustrated in Fig. 6. It was found that the EG on Si-face was primarily monolayer EG except near SiC step edges where bilayer EG was indicated (12). To further validate Raman thickness measurements of EG, cross-sectional TEM was performed. Transmission electron micrographs include a transitional layer (Layer 0), which is in direct contact with the SiC substrate, and generally is not considered graphene. The subsequent layers above Layer 0 constitute the electrically active graphene, and give rise to its unique properties. These first two upper layers are considered monolayer, Fig. 6b and bilayer, Fig. 6c, EG respectively (the arrows point to the transitional layer and the EG). Analysis of Raman measurements of EG on Si-face samples implied the films were primarily monolayer, suggesting that *in vacuo* growth on the Si-face of SiC wafers would result in EG with good thickness uniformity.

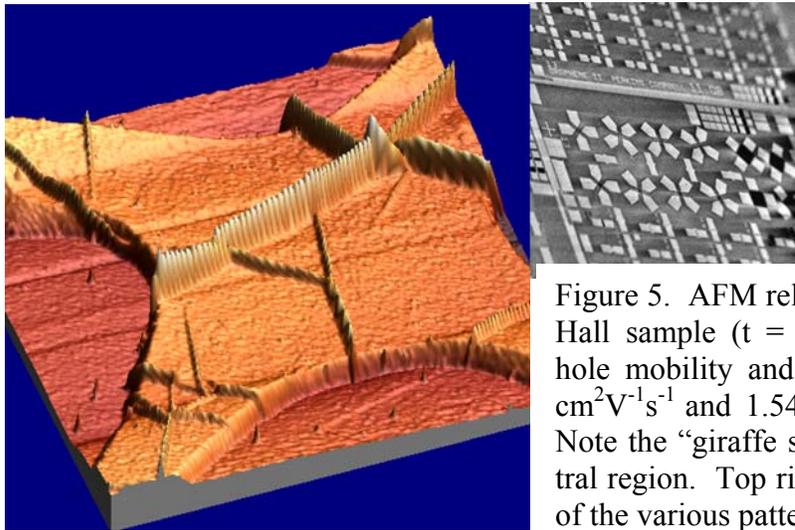

Figure 5. AFM relief map of a 2×2 μm² EG Hall sample (t = 11.7 nm) having 300 K hole mobility and concentration of 11,600 cm²V⁻¹s⁻¹ and 1.54×10¹³ cm⁻², respectively. Note the "giraffe stripe" present in the central region. Top right shows an SEM image of the various patterns placed on the sample.

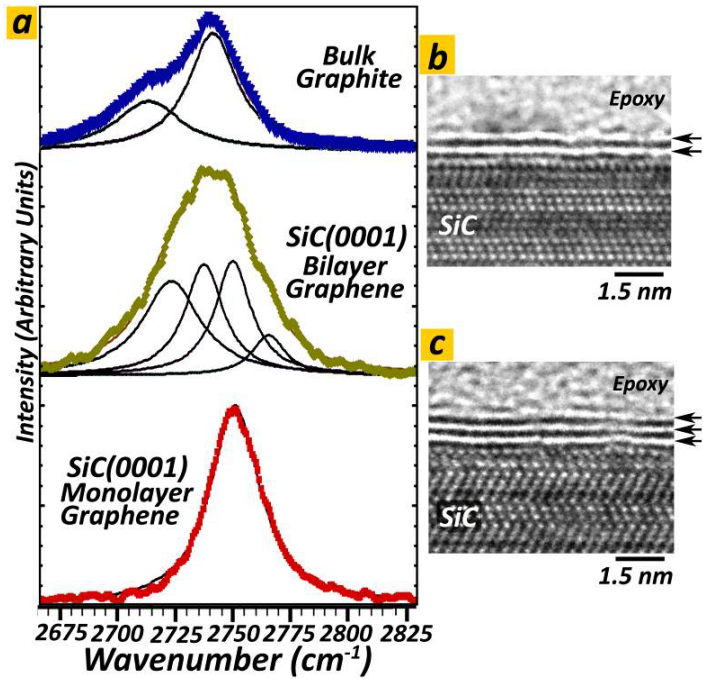

Figure 6. (a) Examples of Raman spectra fits to 2, 4, and 1 Lorentzian(s) corresponding to bulk graphite, and bilayer and monolayer EG, respectively. (b), TEM cross section for monolayer EG on Si-face 6H-SiC and (c) shows a bilayer example. Both micrographs also depict the transition layer (Layer 0) between the EG and substrate. Figure taken from Ref. 14.

Epitaxial growths on 50.8 mm Si-face 6H-SiC wafers were carried out under conditions developed for the smaller area samples; Fig. 7 shows a representative of the first experiments. The morphology of the EG was similar to that of the smaller sized samples described above. After growth, the resistivities of a set of wafers were measured by a non-contact Lehighton probe station and the resistivity uniformity for each wafer was determined. The resistivity uniformity ranged from 11 to 40%. The average Lehighton mobility of the wafer set was also measured, and the averages ranged from 520 to 2780 cm$^2$ V$^{-1}$ s$^{-1}$. This result demonstrates that large area EG grown on Si-face SiC has transport properties similar to small area samples. Maps of strain and thickness were obtained using Raman spectroscopy and showed that most of the EG on each wafer was monolayer. Furthermore, strain was found to be lowest at step edges where the EG was bilayer.

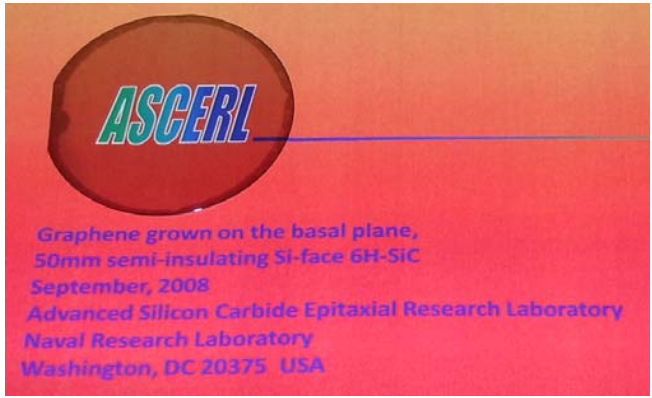

Figure 7. Shown is one of the first 50.8 mm wafers with EG grown using the *in vacuo* approach.

Graphene field effect transistors (FETs) were fabricated with source and drain nonalloyed ohmic metal schemes, where ohmic contact resistances as low as 0.03 Ω·mm were obtained. Metal gates were used on top of an atomic-layer-deposited high-κ (Al$_2$O$_3$) gate dielectric layer with gate lengths of 2 μm. Prototype graphene FETs showed on-state currents as high as 1180 μA μm$^{-1}$ at a drain bias of 1 V and 3000 μA μm$^{-1}$ at 5 V. The $I_{on}/I_{off}$ ratio was 3 to 4. An example FET is shown in Fig. 8. The RF performance, characterized by an HP8510, showed an extrinsic $f_t \cdot L_g$ of 8 GHz·μm with $f_{max}$ of 14 GHz at $V_{ds} = 5$ V. The RF speed performance is expected to be improved as the EG FETs are scaled to gate lengths below 100 nm due to reduced parasitic capacitance and resistance (14).

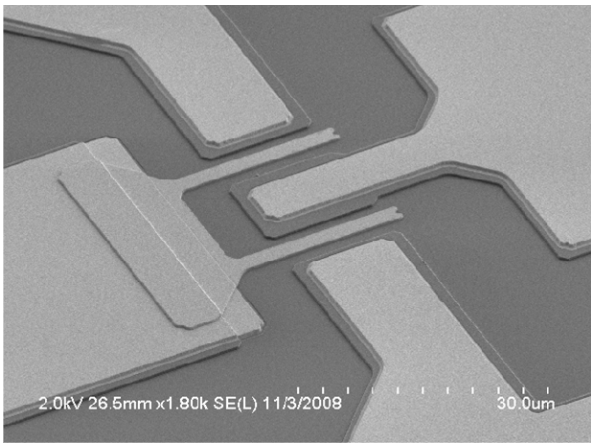

Figure 8. Shown is an example of the first EG FET structure with GHz operation.

## Summary

Epitaxial graphene was grown *in vacuo* by a commercial CVD reactor using a combined hydrogen etch and graphenization process. Growth conditions on Si- and C-face 4H and 6H-SiC semi-insulating samples were investigated and EG properties as a function of growth conditions were determined. Optimal growth conditions were identified and the properties of the EG were found to be similar to the best reports to-date. Characterization by XPS showed that EG on Si-face had a transitional layer whereas C-face did not. Raman spectroscopy showed that the Si-face samples were mainly monolayer EG. These results were used to grow EG on 50.8 mm Si-face 6H-SiC substrates. The morphology of the EG on substrates were similar to those of smaller samples and the electrical properties were similar to or better than those found previously by others. The wafers were processed into FETs and RF performance was evaluated. An $f_{max}$ of 14 GHz at $V_{ds}$ = 5 V was measured - excellent results for these nascent efforts. These results make an encouraging argument that large area EG on SiC is technologically viable and makes an excellent choice for the development of graphene-based device technologies.

## Acknowledgements

This work was supported by the DARPA CERA Program and by the Office of Naval Research. JLT and BLV acknowledge support from the American Society for Engineering Education – Naval Research Laboratory Postdoctoral Fellowship Program.